\documentclass[10pt,conference,review]{IEEEtran}
\usepackage{cite}
\usepackage{amsmath,amssymb,amsfonts}
\usepackage{algorithmic}
\usepackage{graphicx}
\usepackage{textcomp}
\usepackage{xcolor}
\usepackage{url}
\usepackage{hyperref}
\def\BibTeX{{\rm B\kern-.05em{\sc i\kern-.025em b}\kern-.08em
    T\kern-.1667em\lower.7ex\hbox{E}\kern-.125emX}}

\usepackage{booktabs}
\usepackage{url}
\usepackage{listings}
\usepackage{numprint}
\usepackage[inline]{enumitem}
\setlist[enumerate,1]{label=\textit{\alph*)}}
\usepackage{multirow}
\usepackage{caption}
\usepackage{subcaption}

\usepackage{xparse}

\begin{document}

\title{
On the Impact of Language Selection for Training and Evaluating Programming Language Models
}

\author{\IEEEauthorblockN{Jonathan Katzy}
\IEEEauthorblockA{
\textit{Delft University of Technology}\\
Delft, The Netherlands \\
J.B.Katzy@TUDelft.nl}
\and
\IEEEauthorblockN{Maliheh Izadi}
\IEEEauthorblockA{
\textit{Delft University of Technology}\\
Delft, The Netherlands \\
M.Izadi@TUDelft.nl}
\and
\IEEEauthorblockN{Arie van Deursen}
\IEEEauthorblockA{
\textit{Delft University of Technology}\\
Delft, The Netherlands \\
Arie.vanDeursen@TUDelft.nl}
}

\maketitle

\begin{abstract}
    The recent advancements in Transformer-based
Language Models have demonstrated significant potential in
enhancing the multilingual capabilities of these models. The
remarkable progress made in this domain not only applies
to natural language tasks but also extends to the domain of
programming languages. Despite the ability of these models
to learn from multiple languages, evaluations typically focus
on particular combinations of the same languages. In this
study, we evaluate the similarity of programming languages by
analyzing their representations using a CodeBERT-based model.
Our experiments reveal that token representation in languages
such as C++, Python, and Java exhibit proximity to one another,
whereas the same tokens in languages such as Mathematica and
R display significant dissimilarity. Our findings suggest that this
phenomenon can potentially result in performance challenges
when dealing with diverse languages. Thus, we recommend using
our similarity measure to select a diverse set of programming
languages when training and evaluating future models.
\end{abstract}


\section{Introduction}\label{sec:intro}
Language Models (LMs) have shown great capabilities 
in both Natural Language Processing (NLP) and source code domains~\cite{ouyang2022training,chatGPT,bert,feng2020codebert,guo-etal-2022-unixcoder,izadi2022codefill,van2023enriching,alkaswan2023extending}.
Initially, studies focused on the performance of LMs on a single language, 
whereas currently, the emphasis is on 
achieving optimal performance across various languages simultaneously~\cite{chen2021codex, li2023starcoder, xu2022systematic, lu2021codexglue, svyatkovskiy2020intellicode, fried2022incoder, WangCodeT5}. 
The current direction of research suggests that 
pretraining LMs on multiple languages 
benefits per-language performance in many cases~\cite{Chen2022Transferability}.
However, there is a lack of clear rationale or justification 
for the specific selection of languages used in these studies.
Furthermore, research has shown that models 
perform better on certain programming languages and worse on others~\cite{xu2022systematic, svyatkovskiy2020intellicode, li2023starcoder}. 

In the realm of programming, 
source code tokens
are predominantly written in English. These tokens, especially user-defined ones like function and variable names, contain critical information about the code's functionality. Thus, it is reasonable to anticipate that machine learning models should learn from and perform equivalently across different languages. However, significant performance discrepancies persist among LMs working on different languages. This suggests that these performance variations might originate from differing token representations learned by the models.

In this investigation, we aim to answer the following question:
\textbf{How similar are the representations of various programming languages learned by Programming Language Models (PLMs)?}
To answer this question, 
we first identify a set of prevalent languages 
typically employed in the training and evaluation of PLMs, 
as well as a number of frequently disregarded ones in the same context. 
This enables us to include a diverse set of programming languages in our study.
We then leverage a prominent PLM, the CodeBERT model~\cite{feng2020codebert},
to obtain representations of code tokens. 
Next, we calculate language similarity 
by examining the set of tokens shared across all evaluated languages. 
This approach allows us to establish a quantitative basis 
for language selection for PLM research.

Our findings reveal a persistent difference in representations 
across both the multilingual pretrained setting and the non-pretrained monolingual setting.  
To validate our results, we conduct a comparison between the identified similarities and the reported performance of other models. We find that models evaluated on Scala and Ruby perform worse than when they are evaluated on Java, Python, C, C++, Go, and JavaScript~\cite{xu2022systematic}.
This difference in performance overlaps with the relative similarities we have found in these languages.

The implications of this work are twofold.
We have shown that there is an intrinsic difference in the representations of programming languages learned by CodeBERT implying that there is an inherent difference in how PLMs use inputs from different languages.
This finding can have a large impact on areas such as multilingual models and transfer learning.
Theoretically, we have proposed a method for validating language selection in the context of multilingual model outcomes, as well as a foundational understanding to guide language choice for multilingual investigations.
From a practical standpoint, we provide a list of languages with akin representations, indicating similar performance potential. This enables researchers to make informed choices in selecting programming languages aligned with their specific experimental design, especially when aiming to leverage transfer learning techniques across languages.

Finally, we plan to extend this investigation to include more languages to gain a more comprehensive mapping of available languages.
We also want to extend this investigation to other architectures and training goals.
In this work, we used an encoder-based architecture that used the masked language modeling task~\cite{feng2020codebert}, we want to extend the work to other tasks such as code infilling~\cite{fried2022incoder, allal2023santacoder, li2023starcoder} and architectures such as the T5 models~\cite{WangCodeT5}.
Lastly, we release our replication package~\cite{replication}.

\vspace{-2mm}
\section{Approach}\label{sec:approach} 
\begin{table*}[tb]
    \centering
    \scriptsize
    \caption{Selected languages and inclusion criteria}
    \label{tab:language_choices_and_dataset}
    \begin{tabular}{l|l|r|r}
    \toprule
        \textbf{Language} & \textbf{Inclusion criteria} & \textbf{Files} & \textbf{Total Tokens}\\
        \midrule
        Assembly & Unique syntax with a limited vocabulary & 100,000 & 364,776,405\\
        C        & Widely used general-purpose programming language & 100,000 & 326,871,237\\
        COBOL    & The language often present in legacy systems, with a very unique syntax & 2,978 & 10,613,233\\
        C++      & Widely used general-purpose programming language, close to Java and C & 100,000 & 368,090,173\\
        Cuda     & Domain specific application of C++ & 58,355 & 283,624,967\\
        Emacs Lisp & Domain-specific application of Lisp & 54,768 & 188,661,262\\
        Fortran  & Scientific computing language, with similar syntax to Julia and Ruby & 100,000 & 607,478,891\\
        Go       & Domain-specific language with elements from C, C++, Python, and Ruby & 100,000 & 232,054,204\\
        HTML     & Domain-specific language, with unique syntax & 100,000 & 723,969,345\\
        Java     & Widely used general-purpose programming language & 100,000 & 183,040,204\\
        JavaScript & Widely used domain-specific programming language & 100,000 & 325,109,387\\
        Julia    & New emerging scientific computing language & 100,000 & 242,836,338\\
        Kotlin   & Mixture of Java and JS elements but less verbose & 100,000 & 111,578,961\\
        Lisp     & General purpose list-based programming language & 100,000 & 832,184,093\\
        Mathematica & Mathematical computing language with unique features & 26,895 & 1,035,010,885\\
        Python   & General purpose programming language, with semantic whitespace & 100,000 & 237,414,388\\
        R        & Scientific computing language & 39,194 & 154,180,798\\
        Ruby     & General purpose language with syntax similar to Python and Julia & 100,000 & 93,200,451\\
        Scala    & JVM-based language with syntactic elements from JavaScript and C++ & 100,000 & 141,672,916\\
        WebAssembly & Domain-specific emerging list-based language & 5,359 & 59,809,452\\
        \bottomrule
    \end{tabular}
\end{table*}

In our approach, we first train a set of BERT-style models according to the methodology of CodeBERTScore~\cite{zhou2023codebertscore}. 
These models will be used to calculate the representations of tokens. 
Then we calculate the similarity of languages using the tokens that are present in all languages.

\paragraph{Language Selection and the Dataset} 
The training of our models utilizes 
``The Stack"~\cite{Kocetkov2022TheStack}
dataset comprising $358$ programming languages, 
with code availability governed by permissive licenses.

We aim to include a diverse set of languages to encompass as many different scenarios as possible.
There are four main criteria for selection.
First, we look at what languages are commonly used in Machine Learning for Software Engineering research.
Then we look for a variety of grammar and programming paradigms.
We also look at the expected use-case of the languages.
Finally, we look at including both high- and low-resource languages.
From these selection criteria, we curated the list of $20$ languages.
We include the full list of selected languages 
together with the reason for inclusion in Table~\ref{tab:language_choices_and_dataset}.
Note that we do not include some mainstream languages, 
such as C\#, Matlab, and Rust
as their niche within the selection criteria has been sufficiently filled.
Whereas we include some languages that are not often used in the training of such models, 
i.e., Emacs Lisp (ELisp), Lisp, WebAssembly, COBOL, and Fortran because their grammar is very dissimilar to popular languages.
This makes them good subjects to analyze the potential differences in representations.

The dataset that we use for training our models and evaluating the language similarities 
consist of the first $100,000$ files from ``The Stack"~\cite{Kocetkov2022TheStack} for each language.
Note that for low-resource languages where data is scarce (fewer than $100,000$ files), 
we used all available files in The Stack dataset. 
This information is included in Table~\ref{tab:language_choices_and_dataset}.
When training the CodeBERT models, 
we use the first $90\%$ of the files as training data, and the last $10\%$ as a test set.
We calculate the final similarities with representations of samples from the test set.

Note that when selecting common tokens across languages, we exclude comments from consideration. This ensures that we only compare tokens integral to the source code. However, while training the models and inferring the representation of tokens, we do include comments in the input, as they can provide valuable information for the model.

\paragraph{Language Representation}

The individual languages have $47$K, $5$K, and $27$K tokens as maximum, minimum, and average tokens, respectively.
To represent a language, we identify tokens that are present in \textit{all} languages (excluding comments). 
This generates common $2,718$ tokens for which we generate a vector representation denoted as $\mathbf{N}$.
We select common tokens by tokenizing all files, and counting the presence of each token in a language.

We only use tokens that are present in all languages when making a comparison. 
This is because using a token that is not present in a particular language would not allow us to calculate a similarity for that token.

We transform each token 
into a numerical vector that captures its semantic meaning or contextual information.

For every token, we have a set of samples, $\mathbf{M}$. 
Samples correspond to the same token 
but are obtained from different occurrences 
within a given language. 
This enables us to generate the token's representation within a language 
as a set of encodings, denoted as $T$. 
Finally, we represent each language $\mathcal{L}$ 
as a set of the token set $T$ as shown by Equation~\ref{eq:token_language_set}, $l$ is the maximum left context, and $r$ is the maximum right context.
\begin{equation}
    \begin{split}
    T &= \{CodeBERT(t_{n-l}, ... , t^m_{n}, ..., t_{n+r}) | m \in \mathbf{M}\} \\
    \mathcal{L} &= \{T^{\mathcal{L}}_N | N \in 0 ... \mathbf{N}\}
    \end{split}
    \label{eq:token_language_set}
\end{equation}

\paragraph{Language Similarity}
We use a similarity score to quantify the degree of similarity 
between two languages that serve as a measure of their relatedness or resemblance. 
To that end, we employ the cosine similarity between the embeddings of each token in the respective languages. 

Next, we define a similarity function for two sets of vectors. 
This 
allows us to assess the similarity 
between the vector representations of tokens in different languages.
We order a vector pair such that the vectors with the highest cosine similarity are matched up. This will give the highest similarity score possible for the chosen tokens. Once we have the definition of the similarity between two sets of vectors we extend this to the similarity between two languages as shown in Equation~\ref{eq:Vector_set_sim}.
We use the mean to aggregate the similarities of the sets of vectors 
and arrive at a single value that gives the similarity between two languages.

\begin{equation}
    \begin{split}
    sim(t^a_x, T^b) &= max\{cosim(t^a_x, t^b_y) | t^b_y \in T^b\} \\
    sim(T^a, T^b) &= \sum^X_{x=0}\frac{sim(t^a_x, T^b)}{X} \\
    sim(\mathcal{L}^a, \mathcal{L}^b) &= \sum^{\mathbf{N}}_{n= 0}\frac{sim(T^a_n, T^b_n)}{\mathbf{N}}
    \end{split}
    \label{eq:Vector_set_sim}
\end{equation}

The ordering of a vector pair is essential to the collected results.
First, a form of ranking is essential as it allows for the reproduction of the results as the cosine similarity is only defined for two vectors.
Furthermore, using the maximum similarity as an ordering criterion 
gives us an upper bound on the similarities of languages.
This means there is no combination of tokens that would give a higher similarity score,
so all results shown are the best-case scenario for similarity.
Additionally, it minimizes the effect of potential synonyms within the data.
Assuming there are two distinct representations for a given token, this ranking compares the most similar representations between languages.

\paragraph{Model Training}
All the models we train for the experiments have the same architecture, and training setup, i.e., CodeBERT base architecture.\footnote{\url{https://huggingface.co/microsoft/codebert-base-mlm}} 
We train the models in two settings; 
in the first setting, we reinitialize all the weights from scratch for each language. 
In the second setting, 
we use the provided pretrained model as a starting point and finetune it on target languages. 
The pretrained version is trained on Python, Java, JavaScript, PHP, Ruby, and Go.
The models were trained using the CodeBERT training setup~\cite{feng2020codebert}, for 100,000 steps.

\paragraph{Experiments}
To evaluate the similarity of the learned representations by the CodeBERT model, 
we conduct two experimental setups.
In the first experiment, we employ a distance metric to calculate the pairwise distance between all languages. This analysis will enable us to identify the languages that exhibit the greatest similarity in terms of how they represent the same tokens.
Next, we examine the similarity of a language to itself. 
This verifies that the representations are consistent within a language.
We use the same metric as mentioned earlier 
but exclude the token from being compared to itself in Equation~\ref{eq:Vector_set_sim}.
Computing the similarities among a token representation within one language and comparing it to the representation of the same token in another language serves a dual-purpose function.
Firstly, this investigates the consistency of token representations computed by CodebBERT for identical tokens.
Maintaining this internal consistency within a single language is crucial when evaluating its correlations with other languages.
Secondly, this enables us to distinguish whether there exists a subset of highly similar tokens responsible for the entirety of the linguistic correlations, or whether these similarities across languages can be attributed to a more distributed pattern of resemblance among numerous tokens.
Finally, we conduct all mentioned experiments in both a multilingual pretrained setting, as well as a non-pretrained (monolingual) setting to ensure that the differences in representations are consistent.

\section{Results}\label{sec:Results}
Initially, we analyze the outcomes obtained from the distance metric.
Figure~\ref{fig:pairwise_similarity} presents 
pairwise similarity of all languages using the non-pretrained models 
while Figure~\ref{fig:pairwise_similarity_pretrained} 
shows the same metric for the pretrained models. 
In both figures, the languages 
are sorted by average similarity to all other languages.
A darker shade of blue indicates that the languages are more similar. 

\subsection{Cross-lingual Similarity Results}
\begin{figure*}
     \begin{subfigure}[b]{0.5\textwidth}
        \centering
        \includegraphics[width=\textwidth]{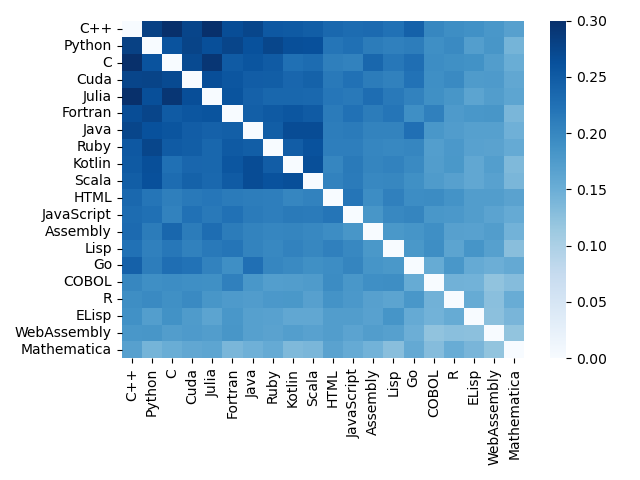}
        \vspace{-4mm}
        \caption{Non-pretrained setting}
        \label{fig:pairwise_similarity}
     \end{subfigure}
    \begin{subfigure}[b]{0.5\textwidth}
        \centering
        \includegraphics[width=\textwidth]{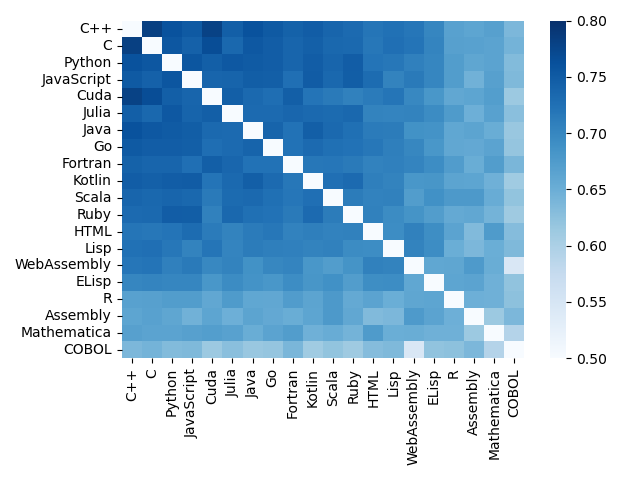}
        \vspace{-4mm}
        \caption{Pretrained setting}
        \label{fig:pairwise_similarity_pretrained}
    \end{subfigure}
    \caption{Pairwise similarity between the representation of all languages}
    \vspace{-4mm}
\end{figure*}
Upon examining the non-pretrained representations of common languages such as C++, Python, Go, Java, and JavaScript, we find that these languages exhibit the closest proximity to each other in terms of their representations. 
Notably, this similarity is observed without any transfer learning from other languages.
When considering derivative languages such as Cuda and Scala, we notice that their representations remain relatively close, but a slight increase in dissimilarity becomes apparent.
When examining the most dissimilar languages, we find that both R and Mathematica exhibit significant differences compared to all other languages.
This observation is intriguing since R and Mathematica are primarily utilized for mathematical and statistical purposes, which distinguishes them from the more commonly used programming languages.
Upon comparison with other languages, we observe that list-based languages, 
namely WebAssembly, Lisp, and ELisp 
exhibit distinct dissimilarities from other languages.
However, they do not demonstrate a significantly higher degree of similarity among themselves 
compared to other languages.
This is a characteristic previously observed in derivative languages like C++ and Cuda.
Additionally, we observe that both COBOL and Assembly (languages with unique syntax), 
also stand significantly apart from all other languages.

For pretrained models, we detect the same trends as previously described. Although the languages, in general, exhibit greater overall similarity, as evident from the change in the scale of Figure~\ref{fig:pairwise_similarity_pretrained}, there remains a discernible difference in the representation of tokens.  
It is worth noting that in the non-pretrained setting, the difference between the highest and lowest similarity score is $0.19$, while in the pretrained setting, this difference is $0.23$. This indicates that although the similarities are closer in the pretrained setting, the differences in similarities remain comparable.
Note that due to the pretraining on languages such as Python, Java, JavaScript, PHP, Ruby, and Go, we observe that these languages become more similar to each other after fine-tuning. However, COBOL, Assembly, ELisp, Lisp, Mathematica, and WebAssembly still maintain their distinctiveness and remain distant from all other languages. 
Another noteworthy observation is that Go, as a result of the pretraining, exhibits increased proximity to a greater number of languages compared to the non-pretrained setting. This indicates that the pretraining process has influenced the language representation, making Go more similar to other languages in the fine-tuned model.

\subsection{Self-Similarity Results}
\begin{figure*}
    \begin{subfigure}[b]{0.48\textwidth}
        \centering
        \includegraphics[width=\textwidth]{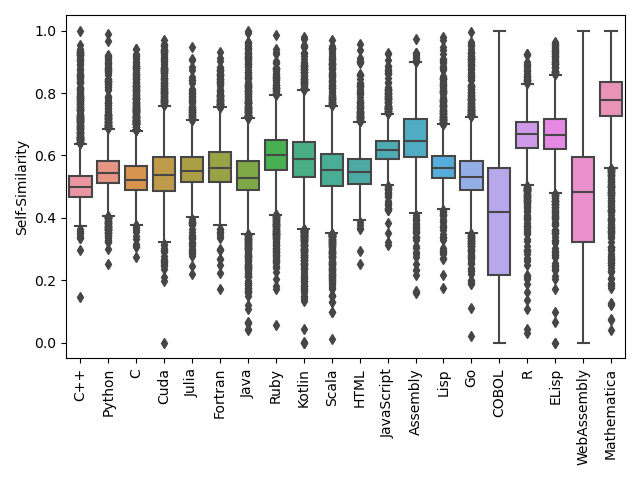}
        \caption{Non-pretrained setting}
        \label{fig:self_sim}
    \end{subfigure}
    \begin{subfigure}[b]{0.48\textwidth}
        \centering
        \includegraphics[width=\textwidth]{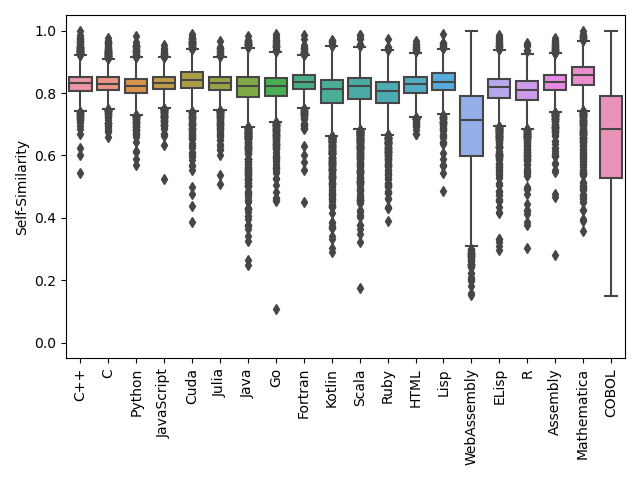}
        \caption{Pretrained setting}
        \label{fig:self_sim_pretrained}
    \end{subfigure}
    \caption{Self-Similarity of languages}
    \vspace{-4mm}
\end{figure*}
Next, we examine the self-similarity of tokens within each language. The self-similarity of non-pretrained representations can be seen in Figure~\ref{fig:self_sim}, 
while the self-similarity of pretrained representations 
is shown in Figure~\ref{fig:self_sim_pretrained}.
Upon analyzing the results, it is evident that when working with a pretrained model, the representation of languages becomes more consistent. 
This observation is supported by two key indicators: the increase in similarity scores and the decrease in variance. Both of these trends reflect the improved consistency and reduced variability in language representations when utilizing a pretrained model.

Our analysis reveals marginally greater token similarity within individual languages compared to inter-language similarities. 
Notably, Mathematica shows the highest self-similarity in both pretrained and non-pretrained settings. 
It also is the least similar to all other languages in the non-pretrained setting, and second least in the pretrained setting. 
These observations emphasize a consistent divergence between Mathematica's representations and those of other languages.

When examining the representations of R, Mathematica, and ELisp, we find that they exhibit consistent patterns similar to other languages in both the monolingual and pretrained settings. This observation confirms the notion that these languages' representations have been learned to be distinct from the more commonly used languages in large language models.

Additionally, we observe that both COBOL and WebAssembly exhibit inconsistency in their representations. We attribute this inconsistency to a potential lack of sufficient training data for these languages, which can result in the models underfitting their representations and failing to capture their inherent linguistic patterns effectively.

The underfitting of two languages provides further insight into inter-language similarities and reinforces the need for meticulous language selection. 
Reviewing non-pretrained model outcomes, Mathematica exhibits greater dissimilarity to other languages compared to both WebAssembly and COBOL, with R and ELisp displaying more dissimilarity than COBOL. 
In the pretrained setting, no language surpasses COBOL in dissimilarity between languages, yet ELisp, R, Assembly, and Mathematica show greater differences compared to WebAssembly. 
This suggests that certain languages produce representations so different from others, that they surpass representations that have not yet converged in dissimilarity.

\section{Discussion}
\subsection{Implications and Future Work}
The implications of our findings lie in the wider field of PLMs. 
We compare the similarities we found between languages to the multilingual performance of a set of PLMs evaluated by Xu et al.~\cite{xu2022systematic} in their study. We see that the trends seen in similarities are mirrored in the performance of the models. Ruby and Scala, the most dissimilar of the languages evaluated by Xu et al. \cite{xu2022systematic} consistently performed worst, while C++, C, Java, Python, and JavaScript had closer and better performances.
Furthermore, we note that the languages used in the evaluation of many models are the same languages which we have found to be the most similar to one another. 
This can have grave implications for the research in this field.

First, when looking at the cross-lingual performance of PLMs, the choice of language will affect the reported performance. Our results help to select representative languages in future studies. 
Second, when investigating the contributions of the language-neutral and language-specific elements of the representations learned by PLMs it will be beneficial to choose a set of languages that are most dissimilar. This should show the greatest difference in the language-specific representations and make the language-neutral elements more evident. 

We have shown that there is an inherent difference in the representations of programming languages for an encoder model trained on the Masked Language Model task. In the future, we aim to extend the investigation to other models such as StarCoder~\cite{li2023starcoder}, or InCoder~\cite{fried2022incoder}, which will use a generative training setup, allowing us to work at a higher granularity than token level, while also including other architectures than only encoder models. Finally, an extension to T5\cite{WangCodeT5} would allow us to evaluate tasks focussed more on code understanding, while maintaining the code generation task also present in StarCoder~\cite{li2023starcoder} and InCoder\cite{fried2022incoder}.

\subsection{Threats to Validity}
Concerning the architecture of the models used, prevalent PLMs generally fall into different categories.
Recent research has indicated the significance of incorporating future context, particularly in models that employ span masking techniques~\cite{fried2022incoder, WangCodeT5, Aghajanyan2022CM3}. In our study, we focus on single token representations, which restricts models that mask spans to operate as BERT-style Masked LMs. 
Concerning the size of the model, using large models is prohibitively expensive as we need to train two models for every language. The current training procedure for $20$ languages took $300$ hours of computing time. This also limits the number of languages analyzed.

Lastly, the selection of the similarity metric potentially affects our findings as well. We employed the cosine similarity, a widely accepted metric in token representations~\cite{word2vec, zhou2023codebertscore}, however, other choices can be investigated as well. 

\section{Related Work}\label{sec:related}
NLP researchers have extensively investigated the multilingual and knowledge transfer performance of encoder models such as Multilingual-BERT~\cite{pires-etal-2019-multilingual}.
As for the similarity measurement, the CodeBERTScore model~\cite{zhou2023codebertscore} 
has demonstrated the potential of using a language model for accuracy assessment and code translations,
proposing a new accuracy metric 
to address limitations in the existing BLEU and ROUGE metrics. 
The authors fine-tune a CodeBERT model~\cite{feng2020codebert} to represent code, enabling the use of the representations as an accuracy metric for code-to-code tasks. Additionally, CodeBERTScore aligns better with human annotators' ratings of code quality, and high-scoring snippets are more likely to be correct programmatically.

\section{Conclusion}\label{sec:conclusion}
In this study, we showed that there is a consistent difference in the representations of the same tokens learned in different languages for BERT-style models. The difference existed in both monolingual and multilingual (finetuned) settings.
We identified several languages that show a high degree of similarity, namely, C, C++, Java, and Python. These languages also correspond to languages commonly used when evaluating code PLMs~\cite{xu2022systematic}.
Furthermore, we specified the languages that learn a consistently different representation for the same tokens, namely, Mathematica and R. This raises inquiries into the performance of PLMs on dissimilar languages, and it calls into question how much these dissimilar languages could benefit from transfer learning.
In future, we will investigate the effects of these differing representations on downstream tasks, as well as investigate the representations found by different model architectures.
\bibliographystyle{IEEEtran}
\bibliography{main}
\end{document}